# The relationship between scientists' research performance and the degree of internationalization of their research[1][2]

*Giovanni Abramo*[a,b,*], *Ciriaco Andrea D'Angelo*[a] *and Marco Solazzi*[a]

[a] Laboratory for Studies of Research and Technology Transfer
School of Engineering, Dept of Management
University of Rome "Tor Vergata"

[b] National Research Council of Italy

**Abstract**

Policy makers, at various levels of governance, generally encourage the development of research collaboration. However the underlying determinants of collaboration are not completely clear. In particular, the literature lacks studies that, taking the individual researcher as the unit of analysis, attempt to understand if and to what extent the researcher's scientific performance might impact on his/her degree of collaboration with foreign colleagues. The current work examines the international collaborations of Italian university researchers for the period 2001-2005, and puts them in relation to each individual's research performance. The results of the investigation, which assumes co-authorship as proxy of research collaboration, show that both research productivity and average quality of output have positive effects on the degree of international collaboration achieved by a scientist.

[1] Abramo, G., D'Angelo, C.A., Solazzi, M. (2011). The relationship between scientists' research performance and the degree of internationalization of their research. *Scientometrics*, 86(3), 629-643. DOI: 10.1007/s11192-010-0284-7

[2] Authors are grateful to Tindaro Cicero for his major support in statistical analysis.

[*] Corresponding author, Dipartimento di Ingegneria dell'Impresa, Università degli Studi di Roma "Tor Vergata", Via del Politecnico 1, 00133 Rome - ITALY, tel. +39 06 72597362, abramo@disp.uniroma2.it

## 1. Introduction

Collaboration in research activity has been the norm for many years[3], particularly in universities, which have the role and mission of sharing knowledge. Collaboration is held as something to be encouraged[4], and over the years there has in fact been a trend towards its increase. Hicks and Katz (1996) examined UK publications in the period 1981-1991, and showed that, not only did the number of co-authors per article increase but also that the number of institutionally co-authored publications increased, both as a number and as a percentage of the total. In 1994, 88% of all UK academic publications involved two or more authors and 55% involved two or more institutions (Katz, 2000). A more recent and more general investigation confirms that the percentage of co-authored publications is increasing over time (Schmoch and Schubert, 2008).

This phenomenon can be linked to a number of factors, including the implementation of specific policies favoring research collaboration, at various levels. At the international level, it is sufficient to note the EU Research Framework Programmes, offering incentives for the research organizations of EU states to carry out cross-national research projects. The increasing costs of research and the complexity of certain undertakings prevent individual researchers, institutions and even nations from taking on certain themes alone (either completely or in an efficient manner). Meanwhile, progress in information and communications technologies, the advent of the internet above all, reduction in transportation costs, and sharpening competition at the local and global levels, with resulting needs for specialization, have all clearly contributed to the increasing adoption of international collaboration in research activity.

The literature on this trend is rich. At the level of world-wide analysis, Zitt and Bassecoulard (2004) examined data from the Science Citation Index (now merged into the Thomson Reuters Web of Science, or WoS) and showed that the percentage of publications realized through international collaboration amounted to less than 10% in 1990, compared to almost 20% of total publications in 2000. Archibugi and Coco (2004) showed that, between 1986 and 1999, "the number of internationally co-authored papers has at least doubled, and in some countries it has tripled". In the case of collaboration involving a specific nation, Schmoch (2005) shows that between 1990 and 2003 the share of German publications co-authored by scientists from other countries grew from 19% to 40%.

The current work inserts in this theme of research on international university collaboration, examining the link between the extent of internationalization of scientific product by a researcher and his or her performance in research. The study is different from those available in the literature, particularly in terms of the unit of analysis: the scientist, rather than the research product. Until now, the positive effect of research collaboration on individual performance has generally been assumed as a given, rather than being tested in any kind of manner (He et al., 2009)[5].

---

[3] "Research collaboration appears to be the rule and not the exception" (Katz, 2000).
[4] "It is widely assumed that collaboration in research is 'a good thing' and that it should be encouraged" (Katz and Martin, 1997).
[5] On one hand, research collaborations are generally undertaken and encouraged because they are viewed as advantageous. On the other hand, and in the direction of causality which this work intends to analyze, the researchers involved in international collaborations could be those with a higher level of performance, who, in virtue of their greater notoriety, experience greater facility in initiating collaborations with

A number of studies based on variously sized sets of publications show that the products originating from research collaborations are characterized by better quality, as measured by citations received or impact factor of the relevant journals. For example, Abramo et al. (2009) report that cross-institution publications by Italian universities for the 2001-2003 triennium were published on average in journals with higher impact factor, when compared to publications from single-institution collaborations. Considering international collaborations, Glänzel and De Lange (2002) analyzed the scientific product of the most active 30 countries in the life sciences in 1995-1996 and show how, although there are differences according to the science field considered, cross-national publications are generally more cited than those realized by researchers from the same nation. A recent study (Suárez-Balseiro et al., 2009), analyzing co-authored articles from Puerto Rico for the years 1980-1999, illustrates that cross-national publications show better positioning in terms of "visibility" (i.e. the rank of the journal where the paper was published in the respective subject area, according to its impact factor), than those originating from both "domestic" cross-institution collaboration and from simple intra-mural collaborations. In the current study we will analyze the link between research performance of a scientist, measured by the bibliometric method, and the degree of internationalization of his or her scientific activity, using co-authorship of scientific publications with foreign authors as a proxy of international collaboration. In particular, after an introductory descriptive analysis of international collaborations, we attempt to answer the following questions:

- Is there a relationship between the degree of international collaboration and the research performance of a university scientist?
- Which of the two indicators of performance, productivity or average quality of scientific product, has the greater impact on intensity and propensity to collaborate at an international level?
- Do the above relationships present differences among the scientific disciplines?

To provide a robust response, the analysis will examine all Italian universities (82) and all the hard sciences, for a total of approximately 26,000 research staff. The analysis will examine scientific product for the period 2001-2005, as censused by the WoS, for a total of over 124,000 publications. The investigation will be conducted at the level of scientific sector in order to avoid distortions due to aggregate measures (Abramo et al., 2008a). All the warnings and cautions pertinent to the limits of the bibliometric approach (van Raan, 2005) apply to the interpretation of the findings of this study.

The next section of this work describes the methodological approach used. Section 3 presents a descriptive analysis of the international collaborations for the population observed. Section 4 reports the results of the investigation. The concluding section comments on the results and indicates possible directions in future research.

## 2. Methodology, dataset and indicators

To analyze the link between performance of individual scientists and the degree of internationalization of their research activity, it is first necessary to characterize the researchers for these two dimensions. The complex nature of the phenomena to be

---

foreign colleagues.

examined requires careful choices: in analytical methodology, type and size of the field of observation, indicators. Each decision involves other lesser choices.

**2.1 Methodological approach**

Research activities resemble a type of input-output process (Moravcsik, 1985), in which the inputs consist of human and financial resources, while outputs have a more complex character, of both tangible (publications, patents, conference presentations, etc.) and intangible nature (personal knowledge, consulting activity, etc.). In terms of output, there are not only many forms for codifying new knowledge, but these are also adopted with differing intensity among the various disciplines. As a particular example, the intensity of publishing research articles in journals varies significantly among subject categories; and so does the coverage of these articles in international bibliometric databases (e.g. Thomson Reuters WoS and Elsevier Scopus). However, limiting the field of investigation to the hard sciences, the literature certainly gives ample justification for the choice of scientific publication as proxy of research output (Moed et al., 2004).

The approach used in the current work is thus decidedly bibliometric, in this case based on the co-authorship of publications in international journals.

The principle limitations of such an approach are that scientific collaborations do not always lead to publication of results and that co-authorship of a publication does not necessarily indicate real collaboration (Melin and Persson 1996; Katz and Martin 1997; Laudel 2002). However, co-authored publication remains one of the most tangible and best documented indicators of research collaboration (Subramanyam 1983; Katz and Martin 1997; Glänzel and Schubert 2004).

**2.2 Data, sources and field of observation**

The data used in this study are obtained from the Observatory on Public Research in Italy (ORP), a bibliometric database developed by the authors. The database is derived from the WoS, and provides a census of the scientific products since 2001 by all research organizations situated in Italy. From the ORP we extracted all publications authored by Italian universities (for the period 2001-2005 there are 147,000 such publications). Through the application of a complex algorithm for the identification of addresses and the disambiguation of the true identity of the authors (see Abramo et al., 2008b for details), it is possible to accurately attribute each publication to the university scientists that produced it[6]. Under the Italian university system, each scientist belongs to

---

[6] For the 147,000 Italian academic publications indexed in the ORP between 2001 and 2005, the statistic harmonic average of precision and recall (F-measure) of authorships as disambiguated by our algorithm is around 95% (sampling error of 2%, confidence level of 98%). When one observes large populations of scientists, the number of homonyms is very high (in the Italian academic system 12% of the 60,000 scientists are affected by homonymy) and their disambiguation within acceptable margins of error is a truly formidable task. According to Zitt and Bassecoulard (2004): "Unification of authors' family names, of institutions such as labs, of lexical forms - in order to avoid synonymy or homonymy lato sensu - can be a bibliometrician's nightmare". This is why bibliometrics-based studies are generally carried out at aggregated levels of analysis, such as university levels. When they are conducted at single scientist level or research group, they are limited to one or few organizations or scientific disciplines. In that case it is

a so-called "scientific disciplinary sector" (SDS). Each SDS in turn forms part of a "university disciplinary area" (UDA). The hard sciences are composed of 9 UDAs[7] and 205 SDSs. The attribution of each publication to its authors and the link between each author and his or her unique SDS permits the attribution of each publication to the SDSs of its authors, and thus the identification of differences that occur between scientific sectors, both in intensity of publication and in collaboration (Abramo et al., 2008a).

To obtain a more reliable representation of the phenomenon under investigation, the analysis is limited to the 165 hard science SDSs in which at least 50% of the member scientists published at least one article in the period under consideration[8]. Further, for greater robustness, the dataset excludes those scientists who entered or left the university system in the period under observation, or who changed SDS or university in this period. It results that there were 26,273 scientists who held a stable faculty position over the observed period, in the 165 SDSs, as indentified from the census of the CINECA database of the Italian Ministry of University and Research[9]. These scientists produced a total of 128,487 publications indexed in the WoS.

To give an idea of the obstacles overcome and the scope of our field of observation, we note the statement by van Raan (2008), one of the world's leading bibliometricians, concerning a dataset consisting of the WoS-listed publications (18,000) of all chemistry scholars (700) in 10 Dutch universities: "This material is quite unique. To our knowledge, no such compilations of very accurately verified publication sets on a large scale are used for statistical analysis of the characteristics of the indicators at the research group level".

## 2.3 Indicators

To examine the link between individual scientists' research performance and the degree of internationalization of their activity, six indicators are defined: three for research performance and three for degree of internationalization.

*Performance indicators*
The research performance of a scientist is measured along two dimensions: productivity and average quality of the research product. The first two of the following indicators concern productivity and the third indicator concerns quality:
- Productivity (P): total of publications authored by a scientist in the period under observation;
- Fractional Productivity (FP): total of the contributions to publications authored by a scientist, with "contribution" defined as the reciprocal of the number of co-authors of each publication;
- Average Quality (AQ): the quality of each publication is proxied by citations of that publication divided by the average number of citations of all publications of the same type (article or review), in the same year and falling in the same subject

---

possible to disambiguate manually. Considering the vast field of observation, the error levels in the dataset used in this study appear more than acceptable.
[7] Mathematics and computer sciences; physics; chemistry; earth sciences; biology; medicine; agricultural and veterinary sciences; civil engineering and architecture; industrial and information engineering.
[8] See Annex for the complete list of the 165 SDS considered.
[9] http://cercauniversita.cineca.it/php5/docenti/cerca.php

category. For example, a value of 1.40 indicates that the publication was cited 40% more often than the average. The average quality equals the sum of standardized citations of all publications by an author divided by the number of his/her publications.

*Indicators of internationalization*

For the other characteristic involved in the analysis, internationalization of research by a scientist, three dimensions are identified: i) intensity, measured by the number of cross-national publications in the period under investigation; ii) propensity, or the ratio of cross-national publications to total publications; iii) amplitude, or number of nations involved in the cross-national publications. The respective indicators are:

- International Collaboration Intensity (ICI): total of publications authored by a scientist in co-authorship with at least one researcher from a foreign organization in the period under observation;
- International Collaboration Rate (ICR): percent ratio of ICI to P;
- International Collaboration Amplitude (ICA): total of foreign nations represented in a cross-national publication by a given scientist.

### 3. General analysis

Before responding to the specific research questions of this study, we provide a descriptive analysis of the degree of internationalization of research activity by Italian scientists, from a geographic and disciplinary viewpoint. Of the 128,487 publications realized by researchers in the dataset over the five years considered, 41,445 were cross-national publications (32.3% of the total). This co-authorship involved 160 foreign nations, as shown in Table 1[10]. The USA is listed first, with 12,560 publications, equal to 30.3% of all Italian cross-national publications. After the USA, the top ten nations collaborating with Italian scientists include Switzerland (sixth position), Russia (eighth) and Canada (ninth), while the remaining top ten are all nations of the European Union, and thus "closer" to Italy, in geo-political terms.

Olmeda Gomez et al. (2009) carried out a similar analysis for Spanish universities, using WoS data for the five-year period 2000-2004. Spanish universities, analogous to those in Italy, realized 30% of their international articles in collaboration with the USA. Further, over 85% of their publications list at least one institution from another EU member state in the WoS "address" field. In the Italian case, this percentage decreases to about 62%.

[Table 1]

Table 2 presents the international collaboration intensity and rate for each UDA[11].

---

[10] Note that the total of column 2 (62,676) is greater than the overall number of publications realized in international co-authorship (41,445), since a publication can be co-authored with more than one foreign nation.

[11] Analogous to the preceding analysis of foreign nations involved, here each publication is counted under each UDA with at least one co-author member. For example, a publication realized by a one researcher in Physics and three in Chemistry is counted (once) for both of these UDAs.

[Table 2]

It can be seen that Medicine, Physics and Biology are the UDAs with the highest number of cross-national publications, while Physics, Earth sciences and Mathematics and computer sciences are those with the highest percentage of cross-national publications out of the total publications for their area.

In the following tables, the level of analysis is deepened to the individual SDSs. Table 3 presents, for each UDA under examination, the SDS with the highest international collaboration rate and the SDS with the lowest ICR, while Table 4 presents the top 10 SDS for ICR, independent of the UDAs to which they belong.

[Table 3]

Both tables illustrate a high international collaboration rate in Physics. In Table 3 it can be seen that for this UDA, the SDS with the lowest ICR (FIS/06[12], at 37.5%) still presents a value higher than all the SDSs of Civil engineering and architecture and of Medicine, and only slightly lower than the maximum registered in Chemistry (CHIM/11, 38%). Note that findings presented in Table 2 and Table 3 are in accordance with those of Abt (2007).

Among the top ten SDSs for ICR (Table 4), four are from Physics (positions 1, 2, 6, 7) and three are from Earth sciences (positions 4, 9, 10). Chemistry, Biology, Medicine and Civil engineering and architecture do not rank any SDSs among the top 10 for ICR.

[Table 4]

Concerning the amplitude of international collaboration, it can be seen that 72.4% of the cross-national publications involve a single foreign nation (Table 5). However, in 17.9% of the cases, the co-authorship involves at least two foreign nations. The cases with involvement of more than five foreign nations represent less than 2% of the total.

[Table 5]

Table 6 presents the data concerning the amplitude of international collaboration, per UDA. Physics is characterized by the lowest percentage of collaborations limited to one foreign nation (63.1%), while Civil engineering and architecture show the highest (85.9%). Medicine shows the highest percentage (7.2%) of publications with more than four foreign nations, while Mathematics and computer sciences show the lowest (0.4%). Considering a number of foreign nations greater than or equal to three, the UDAs showing the highest percentages are Medicine (15.6%) and Physics (13.8%). These same two UDAs also hold the top two positions for average number of foreign nations involved per article: 1.8 for Medicine, 1.6 for Physics. In reality, these areas are characterized by their broad collaboration platforms dedicated to specific large projects, for example in genome and photon studies, involving researchers and organizations from many nations.

The time series for the entire Italian university product, seen in Table 7, confirms the increasing trend for international collaboration. The percentage of cross-national

---

[12] See Annex for the full names of the 165 SDSs that compose the UDAs.

publications has increased in each of the five years under examination, going from 31.4% in 2001 to 33.3% in 2005.
[Table 6]

[Table 7]

Table 8 presents an analysis of the trend at the level of UDA. It is quite clear that the general growth noted in Table 7 is highly conditioned by data from Medicine, the most sizeable area in dimension, with over 40,000 publications realized in the five-year period, or 28% of the total (see Table 2). In fact, in this UDA, the percentage of cross-national publications increased in every year, going from 23.1% in 2001 to 27.1% in 2005.

[Table 8]

**4. Research performance and international collaboration**

This section attempts to provide an answer to the research questions posed in the current study. For this purpose we use the data concerning the six indicators defined in Section 2.3, as measured for each of the 21,504 researchers in the dataset resulting as authors of at least one publication in the WoS over the five-year period under observation (descriptive statistics are presented in Table 9). We will conduct a first analysis of correlation between degree of internationalization of the research activity of specific individuals and their scientific performance, then conduct a detailed analysis of if (and to what extent) the intensity (Section 4.1) and propensity (Section 4.2) for international collaboration of a researcher is influenced by his/her research performance.

[Table 9]

As we would expect (Table 10), the correlation analysis shows a strong link between productivity and international collaboration intensity (Spearman correlation coefficient of 0.653): as the number of publications by an individual scientist increases so does the number of cross-national publications. Similar results are seen for FP, or normalizing the productivity relative to the number of coauthors of the publications: as could be expected, in this case the correlation is slightly lower (0.566). The correlation between ICI and average quality of scientific outputs AQ is also significant (0.380), though not strong.

The correlation between productivity and ICR, while again significant, is quite weak (0.345 for productivity and 0.289 for fractional productivity). The greater or lesser propensity to collaborate with other nations is little correlated to the mass of publications realized. Correlation with average quality is still weaker (0.286).

International collaboration amplitude is strongly correlated to productivity (0.616) and fractional productivity (0.522), as would be reasonable to expect, but much less to average quality (0.379). The same is true for international collaboration intensity. On the other hand, ICI and ICA are strongly correlated: with increasing number of publications authored with foreign colleagues there is also an increase in the number of nations in which these foreign co-authors work. In reality, the value of 0.951 in Table 10 is conditioned by observations of nil value (a researcher that does not collaborate internationally will have nil values for both ICI and ICA). But even excluding such observations from the dataset, the Spearman correlation for ICI and ICA still remains strong (0.761) Further correlation analysis between ICA and the performance indicators would thus be redundant, therefore ICA is excluded from further analysis.

[Table 10]

### 4.1 Research performance and intensity of international collaboration

Various regression analyses are applied to determine if the research performance of a university scientist (independent variable) impacts on international collaborations (dependent variable). First, an attempt is made to determine to what extent the intensity of international collaboration by individual scientists depends on their productivity and on the average quality of their scientific product. Several models are used to attempt to respond to this question: binary logistic, Poisson and negative binomial. Since these provide quite similar results, we present only the results from binary logistic analysis[13], in which the dependent variable ICI is assumed to have a value of one if the scientist has realized at least one cross-national publication, otherwise as nil. The results are presented in Table 11 and Table 12.

[Table 11]
[Table 12]

The good fit of the model is shown by the values of the coefficient of determination (Mc-Fadden's pseudo-R2), the area under the ROC curve, and the percentage of correctly classified. The general statistics and those for single regressors (column 5 of both Tables) show the reliability of the model in representing the link between the observed dependent variable and the independent variables selected. The signs of the coefficients indicate a positive relation between regressors and the dependent variable.

In Table 11, analysis of the standards coefficients shows that P (2.017) has much more weight than AQ (0.353) in determining the probability of collaboration at an international level. For a standard deviation increase in P, the odds of having realized at least one cross national publication are 7.521 times greater, holding the other variable constant; instead for a standard deviation increase in AQ, the odds are 1.424 times greater. With an increase in P from its minimum to maximum Pr(ICI=1) increases by 0.696, while with the same increase in AQ Pr(ICI=1) increases by 0.417.

The same occurs with the model in Table 12, where FP replaces P. Analysis of the standards coefficients shows that P (1.422) has much more weight than AQ (0.491) in determining the probability of collaboration at an international level.

### 4.2 Research performance and rate of international collaboration

The second research hypothesis that we wish to test concerns the possibility that the propensity for international collaboration (represented by ICR) does (or does not) depend on the general research performance of a scientist. Clearly, since the value of the dependent variable (ICR) falls between 0 or 1, we can not use the binary logistic model, as it would give exactly the same results as illustrated in the preceding section. For the same reason, the other models previously noted (Poisson and negative binomial) are not suitable[14]. The ordered logistic regression, through the categorization of the dependent variable, instead permits a discretization and normalization of the values for the dependent variable that, as we will see, assures a good fit of the data for the purposes of the current analysis. In this case, we consider an ordered logistic regression in which the dependent variable is an ordinal category that can take five values:

- 0 if ICR is nil;
- 1 if ICR is between 0.01 and 0.25;
- 2 if ICR is between 0.251 and 0.5;
- 3 if ICR is between 0.501 and 0.75;
- 4 if ICR is between 0.751 and 1.

---

[13] Of the models analyzed, this one presents the lowest value of Akaike IC and highest value of log-likelihood.

The results of the regression are presented in Table 13: Ordered logistic regression of international collaboration rate vs performance indicators (P and AQ).

The international collaboration rate shows significant positive dependence for both P and AQ (Table 13). From the standardized coefficient it can be observed that P (0.448) has slightly greater weight on the dependent variable than does AQ (0.422). For a standard deviation increase in P, the odds of having higher international collaboration rates increase by a factor of 1.565, holding the other variable constant; for a standard deviation increase in AQ, the odds increase by a factor of 1.525.

The positive coefficients of P and AQ indicate an increased probability that a subject with a higher score on either independent variable will be observed in a higher category of ICR. This is confirmed by the results shown in Table 14: increasing P from minimum to maximum value produces an increase in Pr(ICR=4|x) of 0.963, while in category 0 there is a decrease of 0.507. Similar results are also obtained for the variable AQ. However, in the case of FP (Table 15 and Table 16) there is a slight inversion in the results: the weight of AQ (0.462) is slightly greater that that of FP (0.390).

[Table 13]
[Table 14]
[Table 15]
[Table 16]

Side analyses do not show substantial differences among the UDAs: the results shown above continue to hold, with only minimal variations, under specific examination for each UDA.

5. **Conclusions**

The current work takes a bibliometric approach to analyze the link between degree of internationalization of scientific activity and research performance, at the level of individual university researchers.

---

[14] These models suit count data.

To obtain a robust response, the analysis is based on the 124,000 WoS-listed publications over the period 2001-2005 from the 26,000 scientists working in the hard science disciplines of the entire Italian university system.

Collaboration in research is increasing over the years. Our elaborations show that the specific case of international collaboration follows the trend, and that in Italy this is particularly due to events in Medicine, which is the largest disciplinary area. However, in relative terms, Physics shows the highest propensity for international collaboration, with one out of two publications in this area featuring the involvement of foreign authors. Collaborations are often exclusive to two countries: in three quarters of cases, international co-authorships involve only one foreign nation. Arriving at the research questions, the analysis conducted at the level of single researchers shows that the volume of international collaboration is positively correlated to productivity. Such a result appears intuitive: with increasing scientific output by a researcher, there is also an increase of his/her cross-national publications. The correlation between intensity of international collaboration and the average quality of research products by a scientist is not strong. Subsequent regression analyses confirm earlier results.

Productivity has an impact on intensity of international collaboration larger than average quality of scientific product. On the other hand productivity and average quality have a similar weight on the propensity to collaborate with foreign scientists.

The results of the analysis do not change significantly when productivity is standardized for number of coauthors. Further, the results of the general analysis seem to continue to hold when the analysis is conducted at the detailed level of the single disciplinary areas.

However, the results from this aggregate level should be subjected to specific empirical validation. For example, it would be pertinent to examine the situation for specific subpopulations of top scientists, comparing the degree of internationalization of their research activity to that of other colleagues. It would be equally interesting to understand if the difference between quality and productivity in impact on the degree of internationalization varies (or not) depending on the geographic origin of the foreign partner. The authors are currently working on these further questions.

**References**

Abramo G., D'Angelo C.A., Di Costa F., Solazzi M., (2009). University-industry collaboration in Italy: an extensive bibliometric survey. *Technovation*, 29, 498-507.
Abramo G., D'Angelo C.A., Di Costa F., (2008a). Assessment of sectoral aggregation distortion in research productivity measurements. *Research Evaluation*, 17(2), 111-121.
Abramo G., D'Angelo C.A., Pugini F., (2008b). The Measurement of Italian Universities' Research Productivity by a Non Parametric-Bibliometric Methodology. *Scientometrics*, 76(2), 225-244.

Abt H.A., (2007). The frequencies of multinational papers in various sciences. *Scientometrics*, 72(1), 105–115.

Archibugi D., Coco A., (2004). International partnerships for knowledge in business and academia A comparison between Europe and the USA. *Technovation*, 24, 517–528.

Glänzel W., De Lange C., (2002). A distributional approach to multinationality measures of international scientific collaboration. *Scientometrics*, 54, 75–89.

Glänzel W., Schubert A., (2004). Analyzing scientific networks through co-authorship. In: Moed, H. F., Glänzel, W., Schmoch, U. (Eds), Handbook of Quantitative Science and Technology Research: The Use of Publication and Patent Statistics in Studies of S&T Systems, Kluwer, Dordrecht, 257–276.

He Z.L., Geng X.S., Campbell-Hunt C., (2009). Research collaboration and research output: A longitudinal study of 65 biomedical scientists in a New Zealand university. *Research Policy*, 38(2), 306-317.

Hicks D.M., Katz J.S., (1996). Where is science going? *Science, Technology and Human Values*, 21(4), 379–406.

Katz J.S., (2000). Scale independent indicators and research assessment. *Science and Public Policy*, 27(1), 23–36.

Katz J.S., Martin B.R., (1997). What is research collaboration? *Research Policy*, 26, 1–18.

Laudel G., (2002). What do we measure by co-authorships? *Research Evaluation*, 11(1), 3–15.

Melin G., Persson O., (1996). Studying research collaboration using co-authorships. *Scientometrics*, 36(3), 363–367.

Moed H.F., Glänzel W., Schmoch U., (2004). *Handbook of Quantitative Science and Technology Research: The Use of Publication and Patent Statistics in Studies of S & T Systems*. Springer.

Moravcsik M. J., (1985). Applied scientometrics: an assessment methodology for developing countries. *Scientometrics*, 7(3–6), 165–176.

Olmeda Gomez C., Perianes Rodriguez A., Ovalle Perandones M. A., Guerrero Bote V. P., Moya Anegon F. d., (2009). Visualization of scientific co-authorship in Spanish universities: From regionalization to internationalization. *Aslib Proceedings,* 61(1), 83-100.

Schmoch U., Schubert T., (2008). Are international co-publications an indicator for quality of scientific research? *Scientometrics*, 74(3), 361–377.

Schmoch U., (2005). Leistungsfähigkeit und Strukturen der Wissenschaft im Internationalen Vergleich 2004, Bericht zur Technologischen Leistungsfähigkeit, Studien zum Deutschen Innovationssystem Nr. 6-2005, Bundesministerium für Bildung und Forschung.

Suárez-Balseiro C., García-Zorita C., Sanz-Casado E., (2009). Multi-authorship and its impact on the visibility of research from Puerto Rico. *Information Processing and Management*, 45, 469–476.

Subramanyam K., (1983). Bibliometric studies of research collaboration: A review. *Journal of Information Science*, 6(1), 33–38.

van Raan A.F.J., (2005), Fatal attraction: Conceptual and methodological problems in the ranking of universities by bibliometric methods. *Scientometrics*, 62, 133–143.

van Raan A.F.J., (2008). Scaling rules in the science system: Influence of field-specific citation characteristics on the impact of research groups. *Journal of the American Society for Information Science and Technology*, 59(4), 565-576.

Zitt M., Bassecoulard E., (2004). S&T networks and bibliometrics: the case of international scientific collaboration. *4th Proximity Congress: Proximity, Networks and Co-ordination*, Marseille (France), 2004/06/17-18, 15 p.

| Country | Total publications | Incidence (%) in total Italian cross-national publications |
|---|---|---|
| USA | 12,560 | 30.3 |
| France | 6,646 | 16.0 |
| Germany | 5,831 | 14.1 |
| UK | 5,772 | 13.9 |
| Spain | 3,518 | 8.5 |
| Switzerland | 2,438 | 5.9 |
| Netherlands | 2,315 | 5.6 |
| Russia | 1,671 | 4.0 |
| Canada | 1,487 | 3.6 |
| Belgium | 1,458 | 3.5 |
| Other Europe | 11,233 | 27.1 |
| Other Americas | 1,847 | 4.5 |
| Japan | 1,369 | 3.3 |
| China | 555 | 1.3 |
| India | 349 | 0.8 |
| Other Asia | 1,782 | 4.3 |
| Australia | 964 | 2.3 |
| Other Oceania | 166 | 0.4 |
| Africa | 711 | 1.7 |
| Total | 62,676 | |

*Table 1: Classification of foreign nations by number of publications realized in co-authorship with Italian university researchers, 2001-2005*

| | Total publications | | ICI | | ICR (%) | |
|---|---|---|---|---|---|---|
| UDA | Value | Rank | Value | Rank | Value | Rank |
| Mathematics and computer sciences | 11,823 | 6 | 4,043 | 5 | 34.2 | 3 |
| Physics | 18,470 | 4 | 8,887 | 2 | 48.1 | 1 |
| Chemistry | 21,883 | 3 | 7,017 | 4 | 32.1 | 4 |
| Earth sciences | 3,468 | 8 | 1,375 | 8 | 39.6 | 2 |
| Biology | 23,916 | 2 | 7,283 | 3 | 30.5 | 5 |
| Medicine | 40,301 | 1 | 10,077 | 1 | 25.0 | 7 |
| Agricultural and veterinary sciences | 6,641 | 7 | 1,591 | 7 | 24.0 | 9 |
| Civil engineering and architecture | 2,165 | 9 | 581 | 9 | 26.8 | 6 |
| Industrial and information engineering | 15,833 | 5 | 3,837 | 6 | 24.2 | 8 |

*Table 2: International collaboration rate per UDA*

| UDA | SDS with the lowest ICR (%) | | SDS with the highest ICR (%) | |
|---|---|---|---|---|
| Mathematics and computer sciences | INF/01 | 29.5 | MAT/01 | 46.5 |
| Physics | FIS/06 | 37.5 | FIS/05 | 65.2 |
| Chemistry | CHIM/12 | 14.6 | CHIM/11 | 38.0 |
| Earth sciences | GEO/05 | 18.4 | GEO/12 | 52.8 |
| Biology | BIO/03 | 22.8 | BIO/08 | 42.3 |
| Medicine | MED/21 | 9.7 | MED/03 | 36.0 |
| Agricultural and veterinary sciences | VET/10 | 10.6 | AGR/05 | 61.3 |
| Civil engineering and architecture | ICAR/03 | 17.5 | ICAR/02 | 34.6 |
| Industrial and information engineering | ING-IND/17 | 10.6 | ING-IND/20 | 50.8 |

*Table 3: SDSs with the lowest and highest ICR within each UDA*

| SDS | UDA | ICR (%) |
|---|---|---|
| FIS/05 | Physics | 65.2 |
| FIS/04 | Physics | 61.6 |
| AGR/05 | Agricultural and veterinary sciences | 61.3 |
| GEO/12 | Earth sciences | 52.8 |
| ING-IND/20 | Industrial and information engineering | 50.8 |
| FIS/01 | Physics | 47.3 |
| FIS/02 | Physics | 46.9 |
| MAT/01 | Mathematics and computer sciences | 46.5 |
| GEO/01 | Earth sciences | 45.8 |
| GEO/07 | Earth sciences | 45.5 |

*Table 4: Top 10 SDSs for international collaboration rate*

| Number of foreign countries involved | Number of publications | Cross-national publications (% on total) |
|---|---|---|
| 1 | 29,998 | 72.4 |
| 2 | 7,430 | 17.9 |
| 3 | 1,967 | 4.7 |
| 4 | 787 | 1.9 |
| 5 | 452 | 1.1 |
| More than 5 | 149 | 1.8 |

*Table 5: Number of foreign nations involved in cross-national publications*

| UDA | Number of foreign countries involved | | | | | |
|---|---|---|---|---|---|---|
| | 1 | 2 | 3 | 4 | > 4 | Mean |
| Mathematics and computer sciences | 83.2 | 14.0 | 2.2 | 0.3 | 0.4 | 1.2 |
| Physics | 63.1 | 23.1 | 7.5 | 3.0 | 3.3 | 1.6 |
| Chemistry | 78.4 | 16.9 | 3.4 | 0.7 | 0.5 | 1.3 |
| Earth sciences | 73.0 | 18.8 | 3.8 | 1.8 | 2.6 | 1.5 |
| Biology | 75.8 | 17.2 | 4.2 | 1.4 | 1.5 | 1.4 |
| Medicine | 67.7 | 16.7 | 5.5 | 2.9 | 7.2 | 1.8 |
| Agricultural and veterinary sciences | 74.9 | 15.2 | 4.2 | 2.3 | 3.5 | 1.5 |
| Civil engineering and architecture | 85.9 | 11.0 | 1.9 | 0.7 | 0.5 | 1.2 |
| Industrial and information engineering | 81.2 | 13.7 | 3.2 | 0.7 | 1.2 | 1.3 |

*Table 6: Percentage of cross-national publications by number of nations involved: analysis at the level of UDA*

| | 2001 | 2002 | 2003 | 2004 | 2005 | 2001-2005 |
|---|---|---|---|---|---|---|
| Total publications | 22,809 | 24,210 | 26,046 | 27,175 | 28,247 | 128,487 |
| ICI | 7,170 | 7,711 | 8,322 | 8,833 | 9,409 | 41,445 |
| ICR (%) | 31.4 | 31.9 | 32.0 | 32.5 | 33.3 | 32.3 |

*Table 7: ICI and ICR time series*

| UDA | 2001 | 2002 | 2003 | 2004 | 2005 | 2001-2005 |
|---|---|---|---|---|---|---|
| Mathematics and computer sciences | 32.7 | 34.1 | 35.7 | 34.4 | 33.9 | 34.2 |
| Physics | 48.5 | 47.8 | 47.8 | 47.6 | 48.9 | 48.1 |
| Chemistry | 31.3 | 31.1 | 30.5 | 33.3 | 33.9 | 32.1 |
| Earth sciences | 41.0 | 39.7 | 37.6 | 43.7 | 36.5 | 39.6 |
| Biology | 30.5 | 30.2 | 29.6 | 30.2 | 31.6 | 30.4 |
| Medicine | 23.1 | 24.3 | 24.6 | 25.4 | 27.2 | 25.0 |
| Agricultural and veterinary sciences | 25.6 | 26.1 | 22.8 | 22.9 | 23.4 | 24.0 |
| Civil engineering and architecture | 23.3 | 28.7 | 28.9 | 24.2 | 28.4 | 26.8 |
| Industrial and information engineering | 25.1 | 24.0 | 23.5 | 23.4 | 25.2 | 24.2 |

*Table 8: International collaboration rate time series for each UDA*

| Index | Min | Max | Mean | Median | Std. Dev. | Skewness |
|---|---|---|---|---|---|---|
| P | 1 | 247 | 10.20 | 7 | 11.61 | 3.71 |
| FP | 0.02 | 57.8 | 2.20 | 1.43 | 2.52 | 3.98 |
| AQ | 0 | 10.2 | 0.69 | 0.59 | 0.55 | 2.93 |
| ICI | 0 | 167 | 2.71 | 1 | 5.33 | 6.17 |
| ICR | 0 | 1 | 0.22 | 0.12 | 0.27 | 1.27 |
| ICA | 0 | 36 | 1.81 | 1 | 2.77 | 2.91 |

*Table 9: Descriptive statistics of indicators recorded for the dataset (21,504 total observations)*

|  | P | FP | AQ | ICI | ICR | ICA |
|---|---|---|---|---|---|---|
| P | 1 | 0.885* | 0.409** | 0.653** | 0.345** | 0.616** |
| FP |  | 1 | 0.302** | 0.566** | 0.289** | 0.522** |
| AQ |  |  | 1 | 0.380** | 0.286** | 0.379** |
| ICI |  |  |  | 1 | 0.886** | 0.951** |
| ICR |  |  |  |  | 1 | 0.858** |
| ICA |  |  |  |  |  | 1 |

*Table 10: Spearman correlation between indicators used*
*Statistical significance: \*p-value <0.05, \*\*p-value <0.01.*

|  | β | Std Err. | z | P>\|z\| | β$^{sx}$ | e^ β$^{sx}$ | Change of Pr(Y=1) |
|---|---|---|---|---|---|---|---|
| P | 0.174 | 0.003 | 52.34 | 0.000 | 2.017 | 7.521 | 0.696 |
| AQ | 0.640 | 0.033 | 19.42 | 0.000 | 0.353 | 1.424 | 0.417 |
| cons | -1.443 | 0.032 | -45.30 | 0.000 |  |  |  |

*Table 11: Binary logistic regression of international collaboration intensity vs performance indicators (P and AQ).*
*Number of obs: 21,504; LR chi2(2): 6341.16; Prb > chi2: 0.0000; Mc Fadden's R2: 0.2169; Log likelihood: -11455.8; Akaike IC: 22897; Area under ROC (0.8094)*
*β$^{sx}$ = x-standardized coefficient*

|  | β | Std Err. | z | P>\|z\| | β$^{sx}$ | e^ β$^{sx}$ | Change of Pr(Y=1) |
|---|---|---|---|---|---|---|---|
| FP | 0.565 | 0.125 | 45.39 | 0.000 | 1.422 | 4.147 | 0.654 |
| AQ | 0.889 | 0.034 | 26.03 | 0.000 | 0.491 | 1.634 | 0.504 |
| cons | -1.261 | 0.031 | -40.50 | 0.000 |  |  |  |

*Table 12 Binary logistic regression of international collaboration intensity vs performance indicators (FP and AQ).*
*Number of obs: 21,504; LR chi2(2): 4833.42; Prb > chi2: 0.0000; Mc Fadden's R2: 0.1653; Log likelihood: -12199.7; Akaike IC: 24405; Area under ROC (0.7769)*

|  | β | Std Err. | z | P>\|z\| | β$^{sx}$ | e^ β$^{sx}$ |
|---|---|---|---|---|---|---|
| P | 0.039 | 0.001 | 32.93 | 0.000 | 0.448 | 1.565 |
| AQ | 0.764 | 0.026 | 29.33 | 0.000 | 0.422 | 1.525 |
| /cut1 | 0.592 | 0.023 |  |  |  |  |
| /cut2 | 1.729 | 0.026 |  |  |  |  |
| /cut3 | 2.973 | 0.032 |  |  |  |  |
| /cut4 | 3.883 | 0.038 |  |  |  |  |

*Table 13: Ordered logistic regression of international collaboration rate vs performance indicators (P and AQ)*
*Number of obs: 21504; LR chi2(2): 2788.52; Prb > chi2: 0.0000; Mc Fadden's R2: 0.046; Log likelihood: -28628; Akaike IC: 57268*

|        | Change of Pr(Y) |        |        |        |       |
|--------|-----------------|--------|--------|--------|-------|
|        | 0               | 1      | 2      | 3      | 4     |
| **P**  |                 |        |        |        |       |
| Min->Max | -0.507        | -0.255 | -0.155 | -0.046 | 0.963 |
| **AQ** |                 |        |        |        |       |
| Min->Max | -0.549        | -0.241 | -0.134 | -0.033 | 0.957 |

*Table 14: Change in probabilities for categories of international collaboration vs P and AQ*

|       | β     | Std Err. | z     | P>|z| | β$^{sx}$ | e^ β$^{sx}$ |
|-------|-------|----------|-------|-------|----------|-------------|
| FP    | 0.155 | 0.005    | 29.49 | 0.000 | 0.390    | 1.478       |
| AQ    | 0.837 | 0.026    | 32.28 | 0.000 | 0.462    | 1.588       |
| /cut1 | 0.586 | 0.024    |       |       |          |             |
| /cut2 | 1.711 | 0.026    |       |       |          |             |
| /cut3 | 2.951 | 0.032    |       |       |          |             |
| /cut4 | 3.861 | 0.038    |       |       |          |             |

*Table 15: Ordered logistic regression of international collaboration rate vs performance indicators (FP and AQ)*
*Number of obs: 21504; LR chi2(2): 2535.61; Prb > chi2: 0.0000; Mc Fadden's R2: 0.042; Log likelihood: -28754.4; Akaike IC: 57521*

|        | Change of Pr(Y) |        |        |        |       |
|--------|-----------------|--------|--------|--------|-------|
|        | 0               | 1      | 2      | 3      | 4     |
| **FP** |                 |        |        |        |       |
| Min->Max | -0.502        | -0.254 | -0.157 | -0.047 | 0.960 |
| **AQ** |                 |        |        |        |       |
| Min->Max | -0.561        | -0.236 | -0.132 | -0.036 | 0.965 |

*Table 16: Change in probabilities for categories of international collaboration vs FP and AQ*

**Annex - List of SDS considered in the study**

| SDS code | SDS Name | UDA[15] |
|---|---|---|
| INF/01 | Informatics | 1 |
| MAT/01 | Mathematical logic | 1 |
| MAT/02 | Algebra | 1 |
| MAT/03 | Geometry | 1 |
| MAT/05 | Mathematical analysis | 1 |
| MAT/06 | Probability and mathematical statistics | 1 |
| MAT/07 | Mathematical physics | 1 |
| MAT/08 | Numerical analysis | 1 |
| MAT/09 | Operations research | 1 |
| FIS/01 | Experimental physics | 2 |
| FIS/02 | Theoretical physics, mathematical methods and models | 2 |
| FIS/03 | Physics of matter | 2 |
| FIS/04 | Nuclear and subnuclear physics | 2 |
| FIS/05 | Astronomy and astrophysics | 2 |
| FIS/06 | Earth physics and atmospheric environment | 2 |
| FIS/07 | Applied physics (for cultural heritage, environment, biology, medicine) | 2 |
| CHIM/01 | Analytical chemistry | 3 |
| CHIM/02 | Physical chemistry | 3 |
| CHIM/03 | General and inorganic chemistry | 3 |
| CHIM/04 | Industrial chemistry | 3 |
| CHIM/06 | Organic chemistry | 3 |
| CHIM/07 | Fundamentals of chemical technology | 3 |
| CHIM/08 | Pharmaceutical chemistry | 3 |
| CHIM/09 | Pharmaceutical technology | 3 |
| CHIM/10 | Food chemistry | 3 |
| CHIM/11 | Fermentation chemistry and biotechnology | 3 |
| CHIM/12 | Environmental and cultural heritage chemistry | 3 |
| GEO/01 | Paleontology and paleoecology | 4 |
| GEO/02 | Sedimentology and stratigraphy | 4 |
| GEO/03 | Structural geology | 4 |
| GEO/04 | Physical geography and geomorphology | 4 |
| GEO/05 | Applied geology | 4 |
| GEO/06 | Mineralogy | 4 |
| GEO/07 | Petrology and petrography | 4 |
| GEO/08 | Geochemistry and volcanology | 4 |
| GEO/09 | Mining georesources and mineralogical petrographical applications | 4 |
| GEO/10 | Solid earth geophysics | 4 |
| GEO/11 | Applied geophysics | 4 |
| GEO/12 | Oceanography and atmospheric physics | 4 |
| BIO/01 | General botany | 5 |
| BIO/02 | Systematic botany | 5 |
| BIO/03 | Environmental and applied botany | 5 |
| BIO/04 | Plant physiology | 5 |
| BIO/05 | Zoology | 5 |
| BIO/06 | Comparative anatomy and cytology | 5 |
| BIO/07 | Ecology | 5 |
| BIO/08 | Anthropology | 5 |
| BIO/09 | Physiology | 5 |
| BIO/10 | Biochemistry | 5 |
| BIO/11 | Molecular biology | 5 |

---

[15] The UDAs are: 1, Mathematics and computer sciences; 2, Physics; 3, Chemistry; 4, Earth sciences; 5, Biology; 6, Medicine; 7, Agricultural and veterinary sciences; 8, Civil engineering and Architecture; 9, Industrial and information engineering.

| SDS code | SDS Name | UDA[15] |
|----------|----------|---------|
| BIO/12 | Clinical biochemistry and molecular biology | 5 |
| BIO/13 | Applied biology | 5 |
| BIO/14 | Pharmacology | 5 |
| BIO/15 | Pharmaceutical biology | 5 |
| BIO/16 | Human anatomy | 5 |
| BIO/17 | Histology | 5 |
| BIO/18 | Genetics | 5 |
| BIO/19 | General microbiology | 5 |
| MED/01 | Medical statistics | 6 |
| MED/03 | Medical genetics | 6 |
| MED/04 | General pathology | 6 |
| MED/05 | Clinical pathology | 6 |
| MED/06 | Medical oncology | 6 |
| MED/07 | Microbiology and clinical microbiology | 6 |
| MED/08 | Pathological anatomy | 6 |
| MED/09 | Internal medicine | 6 |
| MED/10 | Respiratory diseases | 6 |
| MED/11 | Cardiovascular diseases | 6 |
| MED/12 | Gastroenterology | 6 |
| MED/13 | Endocrinology | 6 |
| MED/14 | Nephrology | 6 |
| MED/15 | Blood diseases | 6 |
| MED/16 | Rheumatology | 6 |
| MED/17 | Infectious diseases | 6 |
| MED/18 | General surgery | 6 |
| MED/19 | Plastic surgery | 6 |
| MED/20 | Pediatric surgery | 6 |
| MED/21 | Thoracic surgery | 6 |
| MED/22 | Vascular surgery | 6 |
| MED/23 | Cardiac surgery | 6 |
| MED/24 | Urology | 6 |
| MED/25 | Psychiatry | 6 |
| MED/26 | Neurology | 6 |
| MED/27 | Neurosurgery | 6 |
| MED/28 | Dentistry | 6 |
| MED/29 | Maxillofacial surgery | 6 |
| MED/30 | Ophthalmology | 6 |
| MED/31 | Otorhinolaryngology | 6 |
| MED/32 | Audiology | 6 |
| MED/33 | Orthopedics | 6 |
| MED/35 | Dermatology | 6 |
| MED/36 | Medical imaging and radiotherapy | 6 |
| MED/37 | Neuroimaging | 6 |
| MED/38 | Pediatrics | 6 |
| MED/39 | Child neuropsychiatry | 6 |
| MED/40 | Obstetrics and gynecology | 6 |
| MED/41 | Anesthesiology | 6 |
| MED/42 | Public, environmental health | 6 |
| MED/44 | Occupational health | 6 |
| AGR/02 | Agronomy and herbaceous crops | 7 |
| AGR/03 | Arboriculture | 7 |
| AGR/05 | Forestry adjustment and forestry | 7 |
| AGR/07 | Agricultural genetics | 7 |
| AGR/08 | Agricultural hydraulic | 7 |
| AGR/11 | General and applied entomology | 7 |
| AGR/12 | Plant pathology | 7 |
| AGR/13 | Agricultural chemistry | 7 |

| SDS code | SDS Name | UDA[15] |
|---|---|---|
| AGR/14 | Pedology | 7 |
| AGR/15 | Food Science and technology | 7 |
| AGR/16 | Agricultural microbiology | 7 |
| AGR/17 | Genetic improvement of livestock | 7 |
| AGR/18 | Animal nutrition | 7 |
| AGR/19 | Special animal husbandry | 7 |
| AGR/20 | Zoocolture | 7 |
| VET/01 | Anatomy of domestic animals | 7 |
| VET/02 | Veterinary physiology | 7 |
| VET/03 | General pathology and veterinary pathological anatomy | 7 |
| VET/04 | Food inspection of animal origin | 7 |
| VET/05 | Infectious diseases of domestic animals | 7 |
| VET/06 | Parasitology and parasitic diseases of animals | 7 |
| VET/07 | Veterinary pharmacology and toxicology | 7 |
| VET/08 | Veterinary medical clinics | 7 |
| VET/09 | Veterinary surgical clinics | 7 |
| VET/10 | Clinical obstetrics and gynecology veterinary | 7 |
| ICAR/01 | Hydraulics | 8 |
| ICAR/02 | Hydrology and hydraulic engineering | 8 |
| ICAR/03 | Environmental health engineering | 8 |
| ICAR/08 | Engineering science building | 8 |
| ICAR/09 | Construction and building technology | 8 |
| ING-IND/03 | Flight mechanics | 9 |
| ING-IND/04 | Aerospace structures | 9 |
| ING-IND/05 | Aerospace equipment and systems | 9 |
| ING-IND/06 | Fluid dynamics | 9 |
| ING-IND/07 | Aerospace propulsion | 9 |
| ING-IND/08 | Fluid machines | 9 |
| ING-IND/09 | Environment and energy systems | 9 |
| ING-IND/10 | Industrial technical physics | 9 |
| ING-IND/12 | Mechanical and thermal measurements | 9 |
| ING-IND/13 | Mechanics applied to machinery | 9 |
| ING-IND/14 | Mechanical design and construction machinery | 9 |
| ING-IND/15 | Engineering drawing | 9 |
| ING-IND/16 | Technologies and processing systems | 9 |
| ING-IND/17 | Industrial plant mechanics | 9 |
| ING-IND/18 | Nuclear reactor physics | 9 |
| ING-IND/19 | Nuclear plant | 9 |
| ING-IND/20 | Measurements and nuclear instrumentation | 9 |
| ING-IND/21 | Metallurgy | 9 |
| ING-IND/22 | Materials science and technology | 9 |
| ING-IND/23 | Applied physical chemistry | 9 |
| ING-IND/24 | Principles of chemical engineering | 9 |
| ING-IND/25 | Chemical plants | 9 |
| ING-IND/26 | Chemical process development | 9 |
| ING-IND/27 | Industrial chemistry and technology | 9 |
| ING-IND/31 | Electrotechnics | 9 |
| ING-IND/32 | Converters, electrical machines and drives | 9 |
| ING-IND/33 | Electrical systems for energy | 9 |
| ING-IND/34 | Bioengineering Industrial | 9 |
| ING-IND/35 | Engineering economics and management | 9 |
| ING-INF/01 | Electronics | 9 |
| ING-INF/02 | Electromagnetic Fields | 9 |
| ING-INF/03 | Telecommunications | 9 |
| ING-INF/04 | Automatics | 9 |
| ING-INF/05 | Information processing systems | 9 |
| ING-INF/06 | Electronic bioengineering and computer science | 9 |

| SDS code | SDS Name | UDA[15] |
|---|---|---|
| ING-INF/07 | Electrical and electronic measurement | 9 |